\documentstyle[12pt]{article}

\newcommand {\nn}    {\nonumber}
\newcommand {\vs}[1]  { \vspace*{#1 cm} }
\newcounter{eq}
\newcounter{sc}

\newcommand {\LMP}  {Lett.Math.Phys.}
\newcommand {\MPL}  {Mod.Phys.Lett.}
\newcommand {\NP}   {Nucl.Phys.}
\newcommand {\PL}   {Phys.Lett.}
\newcommand {\PR}   {Phys.Rev.}
\newcommand {\PRL}   {Phys.Rev.Lett.}
\newcommand {\CMP}  {Comm.Math.Phys.}
\newcommand {\AP}   {Ann.of Phys.}
\newcommand {\PTP}  {Prog.Theor.Phys.}

\newcommand {\CQG}  {Class.Quantum.Grav.}
\def\overleftrightarrow#1{\vbox{\ialign{##\crcr
 $\leftrightarrow$\crcr\noalign{\kern-1pt\nointerlineskip}
 $\hfil\displaystyle{#1}\hfil$\crcr}}}

\newlength{\minitwocolumn}
\begin{document}

%%%%%%%%%%%%%%%%%%%%%%%%%%%%%%%%%%%%%%%%%%%%%%%%%%%%%%%%%%%%%%%%%%
%%%%%%%%%%%%%%%%%%%%%%%% Title %%%%%%%%%%%%%%%%%%%%%%%%%%%%%%%%%%%
%%%%%%%%%%%%%%%%%%%%%%%%%%%%%%%%%%%%%%%%%%%%%%%%%%%%%%%%%%%%%%%%%%

\begin{flushright}
EDO-EP-20\\
June, 1998\\
\end{flushright}
\vspace{15pt}

%\magnification=\magstep1
\pagestyle{empty}
\baselineskip15pt
%\font\cmssB=cmss17
%\font\cmssS=cmss10

\begin{center}
{\large\bf Topological Symmetry, Background Independence \\
and  Matrix Models 
          \footnote{
Invited paper to appear in the special issue of the Journal of
{\it Chaos, Solitons and Fractals} on "Superstrings, M, F, S,
$\ldots$ Theory", edited by M.S. El Naschie and C. Castro.
                  }
\\
\vskip 1mm
}

\vspace{10mm}

Ichiro Oda
          \footnote{
          E-mail address:\ ioda@edogawa-u.ac.jp
                  }
          \footnote{
          Supported in part by Grant-Aid for Scientific Research 
          from Ministry of Education, Science and
          Culture No.09740212.
                  }
\\
\vspace{10mm}
          Edogawa University,
          474 Komaki, Nagareyama City, Chiba 270-01, JAPAN \\

\end{center}

%\maketitle

\vspace{10mm}
\begin{abstract}
We illustrate a physical situation in which topological symmetry,
its breakdown, space-time uncertainty principle, and background 
independence may play an important role in constructing and
understanding matrix models. 
First, we show that the space-time uncertainty principle of string 
may be understood as a manifestation of the breakdown of the 
topological symmetry in the large $N$ matrix model. 
Next, we construct a new type of matrix models which is
a matrix model analog of the topological Chern-Simons and BF 
theories. It is of interest that these topological 
matrix models are not only completely independent of the background 
metric but also have nontrivial "p-brane" solutions as well as 
commuting classical space-time as the classical solutions.
In this paper, we would like to point out some elementary and unsolved
problems associated to the matrix models, whose resolution would lead
to the more satisfying matrix model in future.

\vspace{10mm}

\end{abstract}

\newpage
\pagestyle{plain}
\pagenumbering{arabic}
%\setcounter{page}{1}

%%%%%%%%%%%%%%%%%%%%%%%%%%%%%%%%%%%%%%%%%%%%%%%%%%%%%%%%%%%%%%%%%%
%%%%%%%%%%%%%%%%%%%%%%%% Article %%%%%%%%%%%%%%%%%%%%%%%%%%%%%%%%%
%%%%%%%%%%%%%%%%%%%%%%%%%%%%%%%%%%%%%%%%%%%%%%%%%%%%%%%%%%%%%%%%%%

\rm
%%%%%%%%%%%%%%%%%%%%%%%%%%%%%%%%%%%%%%%%%%%%%%%%%%%%%%%%%%%%%%%%%%%%%
%%%%%%%%%%%%%%%%%%%%%%%%%%%%%%   SEC  1    %%%%%%%%%%%%%%%%%%%%%%%%%%
%%%%%%%%%%%%%%%%%%%%%%%%%%%%%%%%%%%%%%%%%%%%%%%%%%%%%%%%%%%%%%%%%%%%%
\section{Introduction}

It seems that we are now in a new era of developments of quantum
field theories since recent discovery of various remarkable ideas and 
tractable techniques for understanding the strong coupling and the 
non-perturbative regimes of the quantum field theories.
So far the study of quantum field theories has been mainly restricted
to the standard perturbative analyses of weakly coupled field theories.
But in the last few years important progress was made in the study of
the strongly coupled dynamics in a class of gauge theories and string
theory. Such a progress is quite impressive in that
the physics dealing with the strong coupling phase and the 
non-perturbative regime is certainly expected to provide a totally 
new insight about what would be the content of the strong coupling phase
where the conventional perturbative analyses are out of control.

For instance, the discovery of three types of dualities, what we now
call, S, T and U dualities has recently made it possible to clarify 
that five superstring theories and M-theory are in fact 
non-perturbatively equivalent in the sense that each of them 
is nothing but a perturbative expansion of a single underlying theory
about a distinct point in the space of quantum vacua \cite{Schwarz}. 
\footnote{ See the reference \cite{Sugamoto} for old
           fashioned dualities. }
As a second example, we can also list recent progress on the understanding 
of the phase structure of supersymmetric gauge theories in terms of rather 
simple properties of M 5-brane in eleven dimensions \cite{Giveon}. 
In this respect, it is quite interesting that the results of the strongly 
coupled gauge theory are also best understood as string theory 
and M-theory phenomena. Moreover, the more recent conjecture
\cite{Maldacena} that Type IIB superstring on $AdS_5 \times S^5$ is 
equivalent to $N=4$ super Yang-Mills theory with gauge group $SU(N)$
gives rise to a great deal of interests in the study of physics 
on the anti-de Sitter space. This is because according to the above
conjecture the strong coupling regime of $N=4$ super Yang-Mills theory 
with gauge group $SU(N)$ should be described in terms of the weak 
coupling physics of Type IIB superstring on $AdS_5 \times S^5$. 

On the other hand, another striking feature of recent developments 
is a fruitful interplay between superstring theory and a quantum 
theory of black holes. 
It is physically reasonable to imagine that 
a black hole plays a critical role in superstring theory since two 
concepts of black holes and elementary particles would merge at the 
Planck mass scale \cite{'t Hooft} which is relevant to superstring
theory. 
In other words, we expect that black holes may play a role similar
to the hydrogen atom in quantum mechanics in the search of a quantum
theory containing gravity. Referring to a connection with the strong 
coupling physics, black hole quantum mechanics
provides us with a window into strong coupling quantum physics by raising
several puzzles to which a quantum theory of gravity must answer.
This is because in the region of parameter space where elementary particles
become black holes, we inevitably go into strong coupling region.

Among many of the recent remarkable developments in quantum
field theories maybe the most exciting one might be the discovery of matrix 
models.
It is expected that the matrix models may be candidates for the 
non-perturbative formuation of M-theory \cite{M} and 
IIB superstring \cite{IKKT, FKKT} so that a priori they may determine
uniquely the true vacuum among many perturbative vacua of superstring
theory. 
However, it is a pity that although the matrix models have surprisingly 
passed a lot of nontrivial tests to date, it is far from being complete
and further work is being carried out in a number of directions.
For example, it is unclear 
how they yield our real four dimensional space-time and the plausible 
gauge group e.t.c. in the low energy region through some
natural compactification mechanism.
          \footnote{ See the reference \cite{Aoki} for recent
                     development to this problem.}
In such a situation, one of interesting directions of study is to 
ask ourselves whether the matrix models at hand in fact equip with desirable 
characteristic features in themselves as Theory of Everything.
In this context, we would like to ask the following elementary
questions which have not been understood so well, but
it is worthwhile to keep in mind that they should be resolved
to get a more satisfactory matrix model in future:
\begin{itemize}
\item What are the fundamental principles behind matrix theories?
\item What are the underlying gauge symmetries?
\item Is it possible to construct matrix models which do not
depend on the background fields?
\item Why do matrix models involve gravity? In other words,
why is gravity induced from gauge theory?
\item What is a possible mechanism to realize four dimensional
flat space-time?
\end{itemize}

Let us explain the above questions in order in more detail. In order to
explain the first and second questions about the fundamental principles
and the underlying gauge symmetries behind the matrix models,
it is useful to compare the present status of the matrix models
with general relativity by Einstein \cite{Ein}.
General relativity is built from only two basic concepts, 
namely, equivalence principle as the fundamental principle 
and general coordinate invariance as the gauge symmetry
in the framework of Riemannian geometry. These basic concepts
have played a very important role not only in establishing a complete
form of general relativity but also in providing a unified picture
of gravity and geometry.
On the other hand, in the matrix models, we have not yet succeeded in
finding such basic concepts. However, from the successful formulation
of the matrix models one may have a glimpse of a hint that the final 
theory may be constructed based on the non-commutative geometry 
\cite{Connes} which essentially describes the uncertainty principle 
of space-time at the Planck scale, which will be also discussed later.

Concerning the third question, it is useful to cite the words of 
Witten \cite{Witten0}:
"Finding the right framework for an intrinsic, background independent 
formulation of string theory is one of the main problems in the
subject, and so far has remained out of reach." "Though gauge
invariant open-string and closed-string field theories are now
known, the problem of background dependence of string field theory
has not been successfully addressed. This problem is
fundamental because it is here that one really has to address the
question of what kind of geometrical object the string represents."
In other words, string theory and M-theory, as a theory including
quantum gravity, should pick up its own space-time background
in a dynamical manner and should not be a priori formulated 
on the basis of a special background field. 
This problem, of course, is 
closely related to the one of how the theory finds a unique
vacuum state which describes our realistic world.
However, the actions of the matrix models \cite{M, IKKT} 
involve the flat
background metric in the kinetic and/or the potential parts,
which is not allowed from the viewpoint of Witten. We will discuss
how to construct the background independent matrix model in section
3. Maybe in theories dealing with quantum gravity the dynamical
background metric may be induced from some quantum effect or through
a mechanism of spontaneous symmetry breakdown of topological symmetry
\cite{Thorn}. This issue will be also argued in this paper.  

Next, we would like to comment on the fourth question of the relation
between the matrix models and general relativity. 
We should notice that we have at present no clear understanding of
how the matrix theories are connected with Einstein's general
relativity. 
Even if there is circumstancial evidence that the low energy theory
of the matrix models contains general relativity (or supergravity),
it is quite obscure how general relativity is derived from the
matrix models in a comprehensive manner \cite{Bigatti}. 
But the recent progress on $AdS/CFT$ correspondence \cite{Maldacena}
seems to suggest that the supergravity in a bulk theory could be
described by the corresponding gauge theory on its boundary.
Such a viewpoint is physically plausible from the following two
reasons. One reason comes from the Bekenstein-Hawking entropy
formula of black holes, which strongly suggests that if quantum gravity
couples to the theory one space dimension is effectively reduced
and only the boundary theory is relevant to the physics. 
The other reason is that as is well known all observables in
quantum gravity such as mass and charge e.t.c. are defined
on the boundary (usually, at spatial infinity).

Finally, let us consider the problem of Kaluza-Klein compactification
in the matrix models.
Since we wish to regard the matrix models as promising candidates
of the final theory, they should provide a natural mechanism for
compactification yielding the four dimensional flat space-time
from eleven or ten dimensional space-time. With respect to this
point, I have a conjecture although this conjecture is not 
limited to the matrix 
models but connected with a universal feature of string theory
and M-theory. For sake of simplicity, we confine ourselves to the 
eleven dimensional M-theory. M-theory is usually defined as the theory
which possesses $M2$-brane and its EM-dual, $M5$-brane as its
classical solutions. Suppose that $M2$-brane and $M5$-brane occupy
the directions of space-time coordinates along $(t, x_4, x_5)$
and $(t, x_6, x_7, x_8, x_9, x_{10})$, respectively. Here the
key observation is to recall that all extended objects except
string are unstable quantum mechanically as well as classically.
The reason is that roughly speaking we cannot balance
the gravitational attraction with the centrifugal
repulsion in all directions on $p$-branes ($p \ge 2$). 
Then it is natural to make a guess
that the instability associated with $M2$-brane and $M5$-brane
in M-theory plays an important role in the spontaneous compactification
of the excessive space-time dimensions. If we assume that the
space directions tangential to $M2$-brane and $M5$-brane are 
compactified as a result of their instability, we are then left with
the four dimensional space-time $x_\mu \ (\mu = 0, 1, 2, 3)$.
This conjecture is quite speculative (in fact, it is based on a 
simple arithmetic $11 - (2 + 5) = 4$), but it seems to be
surprising at least for me that M-theory has a natural seed
for compactification in the form of classical solutions
in its own right according to the above conjecture.

The paper is organized as follows. In section 2, we construct the
Schild action for general bosonic p-brane that is classically
equivalent to the Nambu-Goto action except some singular configurations,
where special attention is paid to the meaning of the constraints 
in the Schild action for string. 
In section 3, we derive a stronger form of the space-time uncertainty 
principle from the topological field theory where the classical action
is trivially zero. 
The key idea here is the breakdown of the topological symmetry
in passing from the continuous field theory to the discrete
matrix model. 
In section 4, we incorporate the spinors in
the above theory and construct a new matrix model. If we require 
this theory to be invariant under $N=2$ supersymmetric 
transformations in ten dimensions, it turns out that this new 
matrix model becomes the IKKT model or the Yoneya model for Type 
IIB superstring. This choice is dependent on the form of a 
classical solution for a scalar function.
In section 5, we study two types of background independent matrix 
model and examine some intriguing problems such as 
its classical solutions and local symmetries. 
Moreover, we incorporate the spinors in
BF matrix model in a background independent way 
and construct a new matrix model with 
BRST-like supersymmetry whose partition function yields
the Casson invariants. 
The final section is devoted to conclusions.

%%%%%%%%%%%%%%%%%%%%%%%%%%%%%%%%%%%%%%%%%%%%%%%%%%%%%%%%%%%%%%%%%%%%%
%%%%%%%%%%%%%%%%%%%%%%%%%%%%%%   SEC  2    %%%%%%%%%%%%%%%%%%%%%%%%%%
%%%%%%%%%%%%%%%%%%%%%%%%%%%%%%%%%%%%%%%%%%%%%%%%%%%%%%%%%%%%%%%%%%%%%
\section{ The Schild action for general p-brane }

In this section, we construct the Schild action \cite{Schild} for
general bosonic p-brane that is equivalent to the Nambu-Goto action
for p-brane and then analyse the structure of the constraints in
the Hamiltonian formalism \cite{Oda0}.

First of all, let us recall the Schild action \cite{Schild} for
bosonic string ($p = 1$), which is of the form
%**   2.1 %%%%%%%%%%%%%%%%%%%%%%%%%%%%%%%%%%%%%%%%%%%%%%%%%%%%%%%%%
\begin{eqnarray}
S_n ^{p=1} = -\frac{1}{n} \int d^2 \xi \ e \ 
\left[\frac{1}{e^n} \left\{ -\frac{1}{2 \lambda_1 ^2} 
\left( \sigma^{\mu_1 \mu_2} \right)^2 \right\}^{\frac{n}{2}} 
+ n - 1 \right],
\label{2.1}
\end{eqnarray}
%%%%%%%%%%%%%%%%%%%%%%%%%%%%%%%%%%%%%%%%%%%%%%%%%%%%%%%%%%%%%%%%%%%
where $e(\xi)$ is a positive definite scalar density defined 
on the string world sheet parametrized by $\xi^0$ and $\xi^1$, 
$\lambda_1 = 2\pi \alpha'$, and $\sigma^{\mu_1 \mu_2}$ is
defined as $\varepsilon^{\alpha_1 \alpha_2} \partial_{\alpha_1}
X^{\mu_1} \partial_{\alpha_2} X^{\mu_2}$. 
Here $X^\mu (\xi)$ $(\mu = 0, 1, \ldots , D-1)$ are space-time 
coordinates and the index $\alpha$ runs over the world sheet
indices 0 and 1. 
Throughout this paper, we assume that
the flat space-time metric takes the form defined as 
$\eta_{\mu\nu} = diag(- + + \cdots +)$.

Then it is quite straightforward to build the Schild action for
general bosonic p-brane by generalizing (\ref{2.1}) \cite{Oda0}.
The concrete expression is given by
%**   2.2 %%%%%%%%%%%%%%%%%%%%%%%%%%%%%%%%%%%%%%%%%%%%%%%%%%%%%%%%%
\begin{eqnarray}
S_n ^p = -\frac{1}{n} \int d^{p+1} \xi \ e \ 
\left[\frac{1}{e^n} \left\{ -\frac{1}{(p+1)! \lambda_p ^2} 
\left( \sigma^{\mu_1 \cdots \mu_{p+1}} \right)^2 \right\}^{\frac{n}{2}} 
+ n - 1 \right],
\label{2.2}
\end{eqnarray}
%%%%%%%%%%%%%%%%%%%%%%%%%%%%%%%%%%%%%%%%%%%%%%%%%%%%%%%%%%%%%%%%%%%
where $\sigma^{\mu_1 \cdots \mu_{p+1}} = \varepsilon^{\alpha_1 \cdots 
\alpha_{p+1}} \partial_{\alpha_1} X^{\mu_1} \cdots 
\partial_{\alpha_{p+1}} X^{\mu_{p+1}}$ and the world volume index 
$\alpha$ now takes the values $0, 1, \cdots, p$. 

In fact, we can demonstrate that (\ref{2.2}) is equivalent to the 
Nambu-Goto action for p-brane as follows. 
Taking the variation with respect to the auxiliary field $e(\xi)$, 
one obtains the constraint
%**   2.3 %%%%%%%%%%%%%%%%%%%%%%%%%%%%%%%%%%%%%%%%%%%%%%%%%%%%%%%%%
\begin{eqnarray}
e(\xi) = \frac{1}{\lambda_p} \sqrt{-\frac{1}{(p+1)!} 
\left( \sigma^{\mu_1 \cdots \mu_{p+1}} \right)^2}.
\label{2.3}
\end{eqnarray}
%%%%%%%%%%%%%%%%%%%%%%%%%%%%%%%%%%%%%%%%%%%%%%%%%%%%%%%%%%%%%%%%%%%
Plugging the constraint (\ref{2.3}) into the Schild action (\ref{2.2}),
one obtains
%**   2.4 %%%%%%%%%%%%%%%%%%%%%%%%%%%%%%%%%%%%%%%%%%%%%%%%%%%%%%%%%
\begin{eqnarray}
S_n ^p &=& -\int d^{p+1} \xi \ e \nn\\
&=& -\frac{1}{\lambda_p} \int d^{p+1} \xi \sqrt{- \det 
\partial_\alpha X^\mu \partial_\beta X_\mu},
\label{2.4}
\end{eqnarray}
%%%%%%%%%%%%%%%%%%%%%%%%%%%%%%%%%%%%%%%%%%%%%%%%%%%%%%%%%%%%%%%%%%%
where the identity 
%**   2.5 %%%%%%%%%%%%%%%%%%%%%%%%%%%%%%%%%%%%%%%%%%%%%%%%%%%%%%%%%
\begin{eqnarray}
\det \partial_\alpha X^\mu \partial_\beta X_\mu
= \frac{1}{(p+1)!} \left( \sigma^{\mu_1 \cdots \mu_{p+1}} \right)^2
\label{2.5}
\end{eqnarray}
%%%%%%%%%%%%%%%%%%%%%%%%%%%%%%%%%%%%%%%%%%%%%%%%%%%%%%%%%%%%%%%%%%%
was used. Hence the Schild action (\ref{2.2}) becomes at least 
classically equivalent to the Nambu-Goto action (\ref{2.4})
except some singular configurations.

In order to understand the constraint (\ref{2.3}) more closely,
it is useful to make use of the Hamiltonian formalism. 
The canonical conjugate momenta to the $X^\mu$ are given by
%**   2.6 %%%%%%%%%%%%%%%%%%%%%%%%%%%%%%%%%%%%%%%%%%%%%%%%%%%%%%%%%
\begin{eqnarray}
P_\mu &=& \frac{1}{e^{n-1}} \frac{1}{p! \lambda_p ^2} 
\left\{ -\frac{1}{(p+1)! \lambda_p ^2} 
\left( \sigma^{\mu_1 \cdots \mu_{p+1}} \right)^2 
\right\}^{\frac{n}{2}-1} \nn\\
& & {} \times \sigma_{\mu \mu_1 \cdots \mu_p}
\varepsilon^{i_1 \cdots i_p} 
\partial_{i_1} X^{\mu_1} \cdots 
\partial_{i_p} X^{\mu_p},
\label{2.6}
\end{eqnarray}
%%%%%%%%%%%%%%%%%%%%%%%%%%%%%%%%%%%%%%%%%%%%%%%%%%%%%%%%%%%%%%%%%%%
where the index $i$ takes the values from 1 to p. From (\ref{2.6}),
it is easy to see that the momenta satisfy the primary constraints
%**   2.7 %%%%%%%%%%%%%%%%%%%%%%%%%%%%%%%%%%%%%%%%%%%%%%%%%%%%%%%%%
\begin{eqnarray}
P_\mu \partial_i X^\mu &=& 0, \\
P^2 + \frac{1}{\lambda_p ^2} \det \partial_i X^\mu 
\partial_j X_\mu &=& 0,
\label{2.7}
\end{eqnarray}
%%%%%%%%%%%%%%%%%%%%%%%%%%%%%%%%%%%%%%%%%%%%%%%%%%%%%%%%%%%%%%%%%%%
where the lapse (Hamiltonian) constraint (\ref{2.7}) is a 
consequence of the constraint (\ref{2.3}) while the shift (momentum) 
constraints (7) come from the definition (\ref{2.6}) trivially.
In this sense, the constraint (\ref{2.3}) encodes all the dynamical
informations of the Schild action for p-brane. 

Finally, it is valuable to point out that in the case of 
string theory the constraint (\ref{2.3}) expresses the space-time 
uncertainty principle of string \cite{Y1} when the Poisson bracket 
is replaced by a commutator in the large $N$ matrix model, and was 
utilized as the first principle for constructing a type 
IIB supersymmetric matrix model \cite{Y2}. (See the other construction
of matrix models on the basis of the Schild action \cite{Zachos,
Soloviev}.)

%%%%%%%%%%%%%%%%%%%%%%%%%%%%%%%%%%%%%%%%%%%%%%%%%%%%%%%%%%%%%%%%%%%%%
%%%%%%%%%%%%%%%%%%%%%%%%%%%%%%   SEC  3    %%%%%%%%%%%%%%%%%%%%%%%%%%
%%%%%%%%%%%%%%%%%%%%%%%%%%%%%%%%%%%%%%%%%%%%%%%%%%%%%%%%%%%%%%%%%%%%%
\section{ A matrix model from space-time uncertainty
principle and breakdown of topological symmetry }

In this section let us construct a bosonic matrix model which 
expresses an essential content of the space-time uncertainty
principle of string \cite{Oda1}.
Let us start by considering a topological theory \cite{Witten1}
where the classical action is trivially zero but has a nontrivial
dependence on the fields $X^\mu(\xi)$ and $e(\xi)$ as follows:
%**   3.1 %%%%%%%%%%%%%%%%%%%%%%%%%%%%%%%%%%%%%%%%%%%%%%%%%%%%%%%%%
\begin{eqnarray}
S_{c} = S_{c}(X^\mu(\xi), e(\xi)) = 0.
\label{3.1}
\end{eqnarray}
%%%%%%%%%%%%%%%%%%%%%%% %%%%%%%%%%%%%%%%%%%%%%%%%%%%%%%%%%%%%%%%%%%%
The BRST transformations corresponding to the topological
symmetry are given by
%**   3.2 %%%%%%%%%%%%%%%%%%%%%%%%%%%%%%%%%%%%%%%%%%%%%%%%%%%%%%%%%
\begin{eqnarray}
\delta_B X^\mu = \alpha^\mu,  \ \delta_B \alpha^\mu = 0, \nn\\
\delta_B e = e \ \eta,  \ \delta_B \eta = 0, \nn\\
\delta_B \bar{c} = b,  \ \delta_B b = 0,
\label{3.2}
\end{eqnarray}
%%%%%%%%%%%%%%%%%%%%%%%%%%%%%%%%%%%%%%%%%%%%%%%%%%%%%%%%%%%%%%%%%%%
where $\psi^\mu$ and $\eta$ are ghosts, and $\bar{c}$ and $b$
are respectively an antighost and an auxiliary field. Note that
these BRST transformations are obviously nilpotent. Also notice
that the BRST transformation $\delta_B e$ shows the character as a
scalar density of $e$.

As always in the analysis of a topological field theory,
the first step is pick up a gauge which describes an
interesting moduli
space. The key idea in this paper, then, is to fix partially the 
topological symmetry corresponding to $\delta_B e$ by the "conformal" 
constraint (\ref{2.3}) in the case of string.
Consequently, the quantum action defined as $S_b = \int d^2 
\xi \ e L_b$ becomes
%**   3.3 %%%%%%%%%%%%%%%%%%%%%%%%%%%%%%%%%%%%%%%%%%%%%%%%%%%%%%%%%
\begin{eqnarray}
L_b &=&  \frac{1}{e} \delta_B \left[ \bar{c} \left\{ e \left(
\frac{1}{2} \left\{ X^\mu, X^\nu \right\}^2 + \lambda^2 \right)
\right\} \right] \nn\\
&=&  b \left( \frac{1}{2} \left\{ X^\mu, 
X^\nu \right\}^2 + \lambda^2 \right)
- \bar{c} \left( \eta \left( -\frac{1}{2} \left\{ X^\mu, 
X^\nu \right\}^2 + \lambda^2 \right) 
+ 2 \left\{ X^\mu, X^\nu \right\} \left\{ X^\mu, \alpha^\nu 
\right\} \right),
\label{3.3}
\end{eqnarray}
%%%%%%%%%%%%%%%%%%%%%%%%%%%%%%%%%%%%%%%%%%%%%%%%%%%%%%%%%%%%%%%%%%%
where the BRST transformations (\ref{3.2}) were used. Here for later
convenience it is useful to redefine the auxiliary field $b$ by 
$b + \bar{c} \ \eta$. Then $L_b$ can be cast into a simpler form
%**   3.4 %%%%%%%%%%%%%%%%%%%%%%%%%%%%%%%%%%%%%%%%%%%%%%%%%%%%%%%%%
\begin{eqnarray}
L_b =   b \left( \frac{1}{2} \left\{ X^\mu, 
X^\nu \right\}^2 + \lambda^2 \right)
- 2 \lambda^2 \bar{c} \ \eta - 2 \bar{c} \left\{ X^\mu, X^\nu \right\} 
\left\{ X^\mu, \alpha^\nu \right\}.
\label{3.4}
\end{eqnarray}
%%%%%%%%%%%%%%%%%%%%%%%%%%%%%%%%%%%%%%%%%%%%%%%%%%%%%%%%%%%%%%%%%%%

What is necessary to obtain a stronger form of the space-time
uncertainty relation is to move to the large $N$ matrix
theory where we have the following
correspondence
%**   3.5 %%%%%%%%%%%%%%%%%%%%%%%%%%%%%%%%%%%%%%%%%%%%%%%%%%%%%%%%%
\begin{eqnarray}
\int d^2 \xi \ e \longleftrightarrow Trace,\nn\\
\int {\it D} e \longleftrightarrow \sum_{n=1}^\infty,
\label{3.5}
\end{eqnarray}
%%%%%%%%%%%%%%%%%%%%%%%%%%%%%%%%%%%%%%%%%%%%%%%%%%%%%%%%%%%%%%%%%%%
where the trace is taken over $SU(n)$ group. These
correspondence can be justified by expanding the
hermitian matices by $SU(n)$ generators in the large
$N$ limit as is reviewed by the reference \cite{F}.

Now in the large $N$ limit, we have
%**   3.6 %%%%%%%%%%%%%%%%%%%%%%%%%%%%%%%%%%%%%%%%%%%%%%%%%%%%%%%%%
\begin{eqnarray}
S_b = Tr \left( b \left( \frac{1}{2} \left[ X^\mu, 
X^\nu \right]^2 + \lambda^2 \right)
- 2 \lambda^2 \bar{c} \ \eta - 2 \bar{c} \left[ X^\mu, X^\nu \right] 
\left[ X^\mu, \alpha^\nu \right] \right).
\label{3.6}
\end{eqnarray}
%%%%%%%%%%%%%%%%%%%%%%%%%%%%%%%%%%%%%%%%%%%%%%%%%%%%%%%%%%%%%%%%%%%
Then the partition function is defined as
%**   3.7 %%%%%%%%%%%%%%%%%%%%%%%%%%%%%%%%%%%%%%%%%%%%%%%%%%%%%%%%%
\begin{eqnarray}
Z &=& \int {\it D}X^\mu {\it D}\alpha^\mu {\it D}e {\it D}\eta 
{\it D}\bar{c} {\it D}b \ e^{- S_b} \nn\\
&=& \sum_{n=1}^\infty \int {\it D}X^\mu {\it D}\alpha^\mu {\it D}\eta 
{\it D}\bar{c} {\it D}b \ e^{- S_b}.
\label{3.7}
\end{eqnarray}
%%%%%%%%%%%%%%%%%%%%%%%%%%%%%%%%%%%%%%%%%%%%%%%%%%%%%%%%%%%%%%%%%%%
At this stage, it is straightforward to perform the path integration
over $\eta$ and $\bar{c}$. Consequently, one obtains
%**   3.8 %%%%%%%%%%%%%%%%%%%%%%%%%%%%%%%%%%%%%%%%%%%%%%%%%%%%%%%%%
\begin{eqnarray}
Z = \sum_{n=1}^\infty \int {\it D}X^\mu {\it D}\alpha^\mu {\it D}b \
e^{- Tr \ b \ \left( \frac{1}{2} \left[ X^\mu, X^\nu \right]^2 
+ \lambda^2 \right)}.
\label{3.8}
\end{eqnarray}
%%%%%%%%%%%%%%%%%%%%%%%%%%%%%%%%%%%%%%%%%%%%%%%%%%%%%%%%%%%%%%%%%%%
In (\ref{3.8}) since the quantum action does not depend on $\alpha^\mu$ 
it is obvious that there remains the gauge symmetry
%**   3.9 %%%%%%%%%%%%%%%%%%%%%%%%%%%%%%%%%%%%%%%%%%%%%%%%%%%%%%%%%
\begin{eqnarray}
\delta \alpha^\mu = \omega^\mu,
\label{3.9}
\end{eqnarray}
%%%%%%%%%%%%%%%%%%%%%%%%%%%%%%%%%%%%%%%%%%%%%%%%%%%%%%%%%%%%%%%%%%%
which is of course nothing but the remaining topological symmetry. 
Now let us factor out this gauge volume or equivalently fix this 
gauge symmetry by the gauge condition $\alpha^\mu = 0$, so that 
the partition function is finally given by  
%**   3.10 %%%%%%%%%%%%%%%%%%%%%%%%%%%%%%%%%%%%%%%%%%%%%%%%%%%%%%%%%
\begin{eqnarray}
Z = \sum_{n=1}^\infty \int {\it D}X^\mu {\it D}b \
e^{- Tr \ b \ \left( \frac{1}{2} \left[ X^\mu, X^\nu \right]^2 
+ \lambda^2 \right)}.
\label{3.10}
\end{eqnarray}
%%%%%%%%%%%%%%%%%%%%%%%%%%%%%%%%%%%%%%%%%%%%%%%%%%%%%%%%%%%%%%%%%%%

It is remarkable that the variation of the action with respect to the 
auxiliary variable $b$ in (\ref{3.10}) gives a stronger form of the 
space-time uncertainty relation and the theory is 
"dynamical" in the sense that
the ghosts have completely been decoupled in (\ref{3.10}). In other 
words, we
have shown how to derive the space-time uncertainty principle from
a topological theory through the breakdown of the topological symmetry
in the large $N$ matrix model. 
Why has the topological theory yielded
the nontrivial "dynamical" theory? The technical reason is very much simple.
In the passage from the continuous theory (\ref{3.4}) to the matrix theory
(\ref{3.6}),
the dynamical degree of freedom associated with $e(\xi)$ was replaced
by the discrete sum over $n$ while the corresponding
BRST partner $\eta$ remains the continuous variable. This distinct
treatment of the BRST doublet leads to the breakdown of the topological
symmetry giving rise to a "dynamical" matrix theory. In this respect,
it is worthwhile to point out that while the topological symmetry
is "spontaneously" broken in this process, the other gauge symmetries 
never be violated.
Moreover, notice that the above-examined phenomenon is a peculiar
feature in the matrix model with the scalar density $e(\xi)$,
which means that an existence of the gravitational degree of
freedom is an essential ingredient since the generators of the
world-sheet reparametrizations, the Virasoro operators, provide
the Ward-identities associated with the target space general
covariance. 

A rather unexpected appearance of the topological field theory 
also seems to be plausible from the following intuitive arguments. 
Suppose that we live in the world where the topological symmetry 
is exactly valid. 
In such a world we have no means of measuring any distance 
owing to lack of the metric tensor field so that there is neither 
concept of distance nor the space-time uncertainty principle. 
But once the topological symmetry which is particularly connected 
with the gravitational degrees of freedom, is spontaneously broken 
by some dynamical mechanism, 
an existence of the dynamical metric together with a string having 
the minimum length would give us both concepts of the distance and 
the space-time uncertainty principle. Our bosonic matrix model 
actually realizes this scenario in a concrete way.

%%%%%%%%%%%%%%%%%%%%%%%%%%%%%%%%%%%%%%%%%%%%%%%%%%%%%%%%%%%%%%%%%%%%%
%%%%%%%%%%%%%%%%%%%%%%%%%%%%%%   SEC  4    %%%%%%%%%%%%%%%%%%%%%%%%%%
%%%%%%%%%%%%%%%%%%%%%%%%%%%%%%%%%%%%%%%%%%%%%%%%%%%%%%%%%%%%%%%%%%%%%
\section{ Supersymmetric matrix models}

Having obtained a bosonic matrix model, we now turn our attention
to a more interesting model, i.e., its generalization to a 
supersymmetric matrix model \cite{Oda2}.
 Actually, non-perturbative
formulations of both M-theory \cite{M} and IIB superstring 
\cite{IKKT, FKKT}
are based on the supersymmetry. Here we should
emphasize that our philosophy in constructing a supersymmetric
matrix model is rather different from that in the bosonic
case in the previous section although we will go along a similar
path of procedure in what follows. Namely, so far by starting with
the topological field theory \cite{Witten1}, we have tried to derive
the space-time uncertainty principle proposed by Yoneya \cite{Y1,Y2}.
In this section, we promote the space-time uncertainty principle to
one of the basic principles for construction of a supersymmetric
matrix model. In other words, 
on the basis of only two basic principles which are
the space-time uncertainty principle of string and the topological 
symmetry, we attempt to construct a new supersymmetric matrix model.
Of course, in the process of the model building, we will furthermore 
demand a strict invariance under the supersymmetric transformation
although the topological symmetry is broken (in some case even the 
space-time uncertainty principle is not explicit) at the final stage.

As a first step for constructing a supersymmetric matrix model, one
has to require the classical action to depend on the Majorana spinor 
field $\psi_\alpha(\xi)$ as well as the bosonic fields $X^\mu(\xi)$ 
and $e(\xi)$
%**   4.1 %%%%%%%%%%%%%%%%%%%%%%%%%%%%%%%%%%%%%%%%%%%%%%%%%%%%%%%%%
\begin{eqnarray}
S_{c} = S_{c}(X^\mu(\xi), \psi_\alpha(\xi), e(\xi)) = 0,
\label{4.1}
\end{eqnarray}
%%%%%%%%%%%%%%%%%%%%%%%%%%%%%%%%%%%%%%%%%%%%%%%%%%%%%%%%%%%%%%%%%%%
where the subscript $\alpha$ stands for spinor index which should
not be confused with the topological ghost $\alpha^\mu(\xi)$ 
corresponding to $X^\mu(\xi)$. The reason why we
consider only the Majorana spinor will be explained later. This
time, in addition to the BRST transformations (\ref{3.2}) one 
has to add the following BRST transformations for fermions:
%**   4.2 %%%%%%%%%%%%%%%%%%%%%%%%%%%%%%%%%%%%%%%%%%%%%%%%%%%%%%%%%
\begin{eqnarray}
\delta_B \psi_\alpha = \beta_\alpha,  \ \delta_B \beta_\alpha = 0.
\label{4.2}
\end{eqnarray}
%%%%%%%%%%%%%%%%%%%%%%%%%%%%%%%%%%%%%%%%%%%%%%%%%%%%%%%%%%%%%%%%%%%

Next let us set up the gauge condition for $\delta_B e$. Instead
of the bosonic case
%**   4.3 %%%%%%%%%%%%%%%%%%%%%%%%%%%%%%%%%%%%%%%%%%%%%%%%%%%%%%%%%
\begin{eqnarray}
\frac{1}{2} \left\{ X^\mu, X^\nu \right\}^2 + \lambda^2 = 0,
\label{4.3}
\end{eqnarray}
%%%%%%%%%%%%%%%%%%%%%%%%%%%%%%%%%%%%%%%%%%%%%%%%%%%%%%%%%%%%%%%%%%%
we shall set up its natural extension involving the spinor field
%**   4.4 %%%%%%%%%%%%%%%%%%%%%%%%%%%%%%%%%%%%%%%%%%%%%%%%%%%%%%%%%
\begin{eqnarray}
\frac{1}{2} \left\{ X^\mu, X^\nu \right\}^2 + \lambda^2 
+ \frac{1}{2} \bar{\psi} \Gamma_\mu \left\{ X^\mu, \psi \right\}
= 0.
\label{4.4}
\end{eqnarray}
%%%%%%%%%%%%%%%%%%%%%%%%%%%%%%%%%%%%%%%%%%%%%%%%%%%%%%%%%%%%%%%%%%%
When transforming to the matrix theory later, this gauge condition
becomes a generalized stronger form of the space-time uncertainty 
principle. Although this generalized form is different from the
original one proposed by Yoneya \cite{Y1,Y2} by the spinor part,
in the ground state they are obviously equivalent so we take the above
gauge condition (\ref{4.4}). Interestingly enough, it will be
shown later that the gauge choice (\ref{4.4}) leads to the same 
theory as Yoneya's one if a suitable solution for the auxiliary 
variable is chosen. Incidentally, the spinor part in
(\ref{4.4}) is adopted from an analogy with the supersymmetric 
Yang-Mills theory.
Thus we have the quantum action $S_q = \int d^2 \xi \ e \left( L_b 
+ L_f \right)$ with the bosonic contribution $L_b$ (\ref{3.3}) and the
fermionic one $L_f$ given by
%**   4.5 %%%%%%%%%%%%%%%%%%%%%%%%%%%%%%%%%%%%%%%%%%%%%%%%%%%%%%%%%
\begin{eqnarray}
L_f &=&  \frac{1}{e} \delta_B \left( \bar{c} \ e \
\frac{1}{2} \bar{\psi} \Gamma_\mu \left\{ X^\mu, \psi \right\} 
\right)  \nn\\
&=&  b \ \frac{1}{2} \bar{\psi} \Gamma_\mu \left\{ X^\mu, \psi \right\}
- \bar{c} \ \frac{1}{2} \left( \bar{\beta} \Gamma_\mu \left\{ X^\mu,
\psi \right\} - \bar{\psi} \Gamma_\mu \left\{ \alpha^\mu, 
\psi \right\} - \bar{\psi} \Gamma_\mu \left\{ X^\mu, \beta \right\}
\right).
\label{4.5}
\end{eqnarray}
%%%%%%%%%%%%%%%%%%%%%%%%%%%%%%%%%%%%%%%%%%%%%%%%%%%%%%%%%%%%%%%%%%%
Here in a similar way to the bosonic case, let us redefine the
auxiliary field $b$ and the ghost $\beta$ by $b + \bar{c} \ \eta$
and $\beta - \frac{1}{2} \psi \ \eta$, respectively. As a result,
$L_b$ is given by (\ref{3.4}), on the other hand, $L_f$ takes the same 
form as (\ref{4.5}). When we rewrite the fermionic part $L_f$ in this
process, we need the famous Majorana identity $\bar{\psi} 
\Gamma_\mu \psi = 0$, for which we have confined ourselves to 
the Majorana spinor in this paper.  

As before, at this stage let us pass to the matrix model. Again
it is straightforward to carry out the path integration over
$\bar{c}$ and $\eta$ in a perfect analogous way to the bosonic
theory. Accordingly, we arrive at the following partition 
function
%**   4.6 %%%%%%%%%%%%%%%%%%%%%%%%%%%%%%%%%%%%%%%%%%%%%%%%%%%%%%%%%
\begin{eqnarray}
Z = \sum_{n=1}^\infty \int {\it D}X^\mu {\it D}\alpha^\mu 
{\it D}\psi_\alpha {\it D}\beta_\alpha {\it D}b \
e^{- Tr \ b \ \left( \frac{1}{2} \left[ X^\mu, X^\nu \right]^2 
+ \lambda^2 + \frac{1}{2} \bar{\psi} \Gamma_\mu \left[ X^\mu, \psi 
\right] \right)}.
\label{4.6}
\end{eqnarray}
%%%%%%%%%%%%%%%%%%%%%%%%%%%%%%%%%%%%%%%%%%%%%%%%%%%%%%%%%%%%%%%%%%%
In this expression since the quantum action is independent of 
$\alpha^\mu$ and $\beta_\alpha$ we have the remaining topological
symmetries given by
%**   4.7 %%%%%%%%%%%%%%%%%%%%%%%%%%%%%%%%%%%%%%%%%%%%%%%%%%%%%%%%%
\begin{eqnarray}
\delta \alpha^\mu = \omega^\mu, \ \delta \beta_\alpha = 
\rho_\alpha.
\label{4.7}
\end{eqnarray}
%%%%%%%%%%%%%%%%%%%%%%%%%%%%%%%%%%%%%%%%%%%%%%%%%%%%%%%%%%%%%%%%%%%
After factoring these gauge volumes out, the partition function is 
finally cast to be 
%**   4.8 %%%%%%%%%%%%%%%%%%%%%%%%%%%%%%%%%%%%%%%%%%%%%%%%%%%%%%%%%
\begin{eqnarray}
Z &=& \sum_{n=1}^\infty \int {\it D}X^\mu {\it D}\psi_\alpha 
{\it D}b \ e^{- S_q}  \nn\\
&=& \sum_{n=1}^\infty \int {\it D}X^\mu {\it D}\psi_\alpha 
{\it D}b \
e^{- Tr \ b \ \left( \frac{1}{2} \left[ X^\mu, X^\nu \right]^2 
+ \lambda^2 + \frac{1}{2} \bar{\psi} \Gamma_\mu \left[ X^\mu, \psi 
\right] \right)}.
\label{4.8}
\end{eqnarray}
%%%%%%%%%%%%%%%%%%%%%%%%%%%%%%%%%%%%%%%%%%%%%%%%%%%%%%%%%%%%%%%%%%%
Of course, the action $S_q$ still possesses the zero volume
reduction of the usual gauge symmetry
%**   4.9 %%%%%%%%%%%%%%%%%%%%%%%%%%%%%%%%%%%%%%%%%%%%%%%%%%%%%%%%%
\begin{eqnarray}
\delta \psi_\alpha &=& i \left[X_\mu, \Lambda \right], \nn\\
\delta X_\mu &=& i \left[\psi, \Lambda \right], \nn\\ 
\delta b &=& i \left[b, \Lambda \right].
\label{4.9}
\end{eqnarray}
%%%%%%%%%%%%%%%%%%%%%%%%%%%%%%%%%%%%%%%%%%%%%%%%%%%%%%%%%%%%%%%%%%%

In this way, we have constructed a new matrix model with the 
Majorana spinor variable on the basis of the space-time uncertainty
principle and the topological symmetry. Although the action
contains the spinor variable in addition to the bosonic
variable, it is not always supersymmetric. The supersymmetry plays 
the most critical role in the matrix models for M-theory
\cite{M} and IIB superstring theory \cite{IKKT}, 
so we should require the invariance under the supersymmetry 
for the action $S_q$ obtained in (\ref{4.8}). The most natural 
form of $N=2$ supersymmetric transformations is motivated
by a supersymmetric Yang-Mills theory whose
(0+0)-dimensional reduction is given by
%**   4.11 %%%%%%%%%%%%%%%%%%%%%%%%%%%%%%%%%%%%%%%%%%%%%%%%%%%%%%%%%
\begin{eqnarray}
\delta \psi_\alpha^{ab} &=& i \left[X_\mu, X_\nu \right]^{ab}
\left(\Gamma^{\mu\nu} \varepsilon \right)_\alpha + \zeta_\alpha
\delta^{ab}, \nn\\
\delta X_\mu^{ab} &=& i \bar{\varepsilon} \Gamma_\mu 
\psi^{ab}, \nn\\
\delta b^{ab} &=& 0,
\label{4.11}
\end{eqnarray}
%%%%%%%%%%%%%%%%%%%%%%%%%%%%%%%%%%%%%%%%%%%%%%%%%%%%%%%%%%%%%%%%%%%
where we have explicitly written down the matrix indices to
clarify that $\varepsilon_\alpha$ and $\zeta_\alpha$ are the 
Majorana spinor parameters. These supersymmetric transformations 
are of the same form as in IKKT model \cite{IKKT}. At this stage,
we assume the space-time dimensions to be ten in order to make contact 
with IIB superstring. 

To make the action $S_q$ in (\ref{4.8}) invariant under the $N=2$ 
supersymmetry (\ref{4.11}), it is easy to check that $b^{ab}$ must 
take the diagonal form with respect to the hermitian matrix indices. 
There are two interesting solutions. One of them is to select the 
auxiliary variable $b^{ab}$ to be proportional to $\delta^{ab}$ up
to a constant. Without generality we take the proportional constant
to be $- \frac{1}{2}$, therefore
%**   4.12 %%%%%%%%%%%%%%%%%%%%%%%%%%%%%%%%%%%%%%%%%%%%%%%%%%%%%%%%%
\begin{eqnarray}
b^{ab} = - \frac{1}{2} \delta^{ab}.
\label{4.12}
\end{eqnarray}
%%%%%%%%%%%%%%%%%%%%%%%%%%%%%%%%%%%%%%%%%%%%%%%%%%%%%%%%%%%%%%%%%%%
Here if we redefine $X^\mu$, $\psi$, and $- \frac{1}{2} \lambda^2$
in terms of $\alpha^{\frac{1}{4}} X^\mu$, $\sqrt{2} 
\alpha^{\frac{3}{8}} \psi$, and $\beta$, respectively, the action 
$S_q$ can be rewritten to be
%**   4.13 %%%%%%%%%%%%%%%%%%%%%%%%%%%%%%%%%%%%%%%%%%%%%%%%%%%%%%%%%
\begin{eqnarray}
S_q = \alpha \left(- \frac{1}{4} Tr \left[ X^\mu, X^\nu \right]^2 
- \frac{1}{2} Tr \bar{\psi} \Gamma_\mu \left[ X^\mu, \psi 
\right] \right) + \beta Tr \mathbf{1}.
\label{4.13}
\end{eqnarray}
%%%%%%%%%%%%%%%%%%%%%%%%%%%%%%%%%%%%%%%%%%%%%%%%%%%%%%%%%%%%%%%%%%%
Note that this action is completely equivalent to the action in
the IKKT model \cite{IKKT}. In this case, we cannot derive the
space-time uncertainty relation from the equation of motion,
but this relation might be encoded implicitly in the matrix 
character of the model.

The other interesting solution would be of the form
%**   4.14 %%%%%%%%%%%%%%%%%%%%%%%%%%%%%%%%%%%%%%%%%%%%%%%%%%%%%%%%%
\begin{eqnarray}
b^{ab} = c \ \delta^{ab},
\label{4.14}
\end{eqnarray}
%%%%%%%%%%%%%%%%%%%%%%%%%%%%%%%%%%%%%%%%%%%%%%%%%%%%%%%%%%%%%%%%%%%
with some additional auxiliary variable $c$. With this choice, the
partition function (\ref{4.8}) can be reduced to be 
%**   4.15 %%%%%%%%%%%%%%%%%%%%%%%%%%%%%%%%%%%%%%%%%%%%%%%%%%%%%%%%%
\begin{eqnarray}
Z &=& \sum_{n=1}^\infty \int {\it D}X^\mu {\it D}\psi_\alpha 
{\it D}c \ e^{- S_q}  \nn\\
&=& \sum_{n=1}^\infty \int {\it D}X^\mu {\it D}\psi_\alpha 
{\it D}c \
e^{- \ c \ Tr \left( \frac{1}{2} \left[ X^\mu, X^\nu \right]^2 
+ \lambda^2 + \frac{1}{2} \bar{\psi} \Gamma_\mu \left[ X^\mu, \psi 
\right] \right)}.
\label{4.15}
\end{eqnarray}
%%%%%%%%%%%%%%%%%%%%%%%%%%%%%%%%%%%%%%%%%%%%%%%%%%%%%%%%%%%%%%%%%%%
At first sight, it seems that we have obtained a new supersymmetric
matrix model, but this is an illusion. We shall show that the above
model is entirely equivalent to the Yoneya model \cite{Y2} in what 
follows. Provided that we take account of the stronger form of the 
space-time uncertainty principle instead of the weaker form, the 
Yoneya model can be expressed in terms of the partition function
%**   4.16 %%%%%%%%%%%%%%%%%%%%%%%%%%%%%%%%%%%%%%%%%%%%%%%%%%%%%%%%%
\begin{eqnarray}
Z &=& \sum_{n=1}^\infty \int {\it D}X^\mu {\it D}\psi_\alpha 
{\it D}c \ e^{- S_y}  \nn\\
&=& \sum_{n=1}^\infty \int {\it D}X^\mu {\it D}\psi_\alpha 
{\it D}c \
e^{- \ c \ Tr \left( \frac{1}{2} \left[ X^\mu, X^\nu \right]^2 
+ \lambda^2 \right) - Tr \frac{1}{2} \bar{\psi} \Gamma_\mu 
\left[ X^\mu, \psi \right] }.
\label{4.16}
\end{eqnarray}
%%%%%%%%%%%%%%%%%%%%%%%%%%%%%%%%%%%%%%%%%%%%%%%%%%%%%%%%%%%%%%%%%%%
This partition in the Yoneya model does not look like the partition
(\ref{4.15}). But Yoneya has defined the
supersymmetric transformations in a slightly different manner
compared to ours (\ref{4.11}). His supersymmetry is
%**   4.17 %%%%%%%%%%%%%%%%%%%%%%%%%%%%%%%%%%%%%%%%%%%%%%%%%%%%%%%%%
\begin{eqnarray}
\delta \psi_\alpha^{ab} &=& i c  \left[X_\mu, X_\nu \right]^{ab}
\left(\Gamma^{\mu\nu} \varepsilon \right)_\alpha + \zeta_\alpha
\delta^{ab}, \nn\\
\delta X_\mu^{ab} &=& i \bar{\varepsilon} \Gamma_\mu 
\psi^{ab}, \nn\\
\delta c &=& 0.
\label{4.17}
\end{eqnarray}
%%%%%%%%%%%%%%%%%%%%%%%%%%%%%%%%%%%%%%%%%%%%%%%%%%%%%%%%%%%%%%%%%%%  
Note that there exists $c$ variable in the first term
of the right-handed side in the first equation while it is 
absent in our formula (\ref{4.11}) (Of course, in (\ref{4.11}) 
we should replace $\delta b^{ab} = 0$ with $\delta c = 0$ 
for present consideration). Then it is easy to show that if
we redefine $\psi$, $\varepsilon$ and $\zeta$ by
$c^{\frac{1}{2}} \psi$, $c^{-\frac{1}{2}} \varepsilon$ and 
$c^{\frac{1}{2}} \zeta$, respectively in the Yoneya model,
his action $S_y$ and supersymmetric transformations
(\ref{4.17}) conform to our action $S_q$ and supersymmetric
transformations (\ref{4.11}), respectively. To demonstrate 
a complete equivalence, we have to consider the functional 
measure. From these redefinitions
the functional measure receives a contribution of an additional 
factor $c^8$, but this change is absorbed into a definition 
of the functional measure ${\it D}c$ since the variable $c$
is the supersymmetrically invariant non-dynamical auxiliary 
variable in the model at hand. 
In this way, we can show that the solution (\ref{4.14})
gives rise to the Yoneya model. It is surprising that depending
on a choice of the scalar function $b$ our model leads to
the IKKT model \cite{IKKT} and the Yoneya model \cite{Y2},
which on reflection clarifies the difference between both 
the matrix models.

Our approach heavily relies on the mechanism of the breakdown
of the topological symmetry, so we should examine more closely
the reason why our model gives rise to the nontrivial 
"dynamical" theory from at least classically trivial
topological theory.  As mentioned in section 3, the technical
reason lies in asymmetric treatment between the BRST doublet
$e$ and $\eta$. However, there exists a deeper reason behind
it. To make our arguments clear, it is useful to compare the 
present approach with the previous studies about the topological
(pregauge-) pregeometric models \cite{Roberto, Akama} whose 
essential ideas will be recapitulated in what follows.

For generality, we consider an arbitrary dimension of space-time. 
We take the Nambu-Goto action as a classical action 
where we restrict ourselves to the case that the dimension 
is equal between world-volume and space-time. 
Then  we can prove that 
this classical action becomes topological because we
can eliminate all the dynamical degrees of freedom by means of
the world-volume reparametrizations.  Let us rewrite it
to the Polyakov form
%**   4.19 %%%%%%%%%%%%%%%%%%%%%%%%%%%%%%%%%%%%%%%%%%%%%%%%%%%%%%%%%
\begin{eqnarray}
S &=& -\frac{1}{\lambda} \int d^D \xi \sqrt{- \det 
\partial_a X \cdot \partial_b X} \nn\\
&=& \int d^D \xi \sqrt{-g} \left( g^{ab} \partial_a X^\mu
\partial_b X^\nu + \lambda \right).
\label{4.19}
\end{eqnarray}
%%%%%%%%%%%%%%%%%%%%%%%%%%%%%%%%%%%%%%%%%%%%%%%%%%%%%%%%%%%%%%%%%%%
In spite of lack of proof, the above two actions might be 
equivalent even in the quantum level 
owing to the topological character where there is no
anomaly. Next work is to evaluate the effective action for
the metric $g^{ab}$ due to the quantum fluctuation of the
"matter" fields $X^\mu$ whose result is given by \cite{Tera}
%**   4.20 %%%%%%%%%%%%%%%%%%%%%%%%%%%%%%%%%%%%%%%%%%%%%%%%%%%%%%%%%
\begin{eqnarray}
S_{eff} = i \ Tr \log \left[\left( \partial_a \sqrt{-g} g^{ab}
\partial_b \right)\right] +  \lambda \int d^D \xi \sqrt{-g}.
\label{4.20}
\end{eqnarray}
%%%%%%%%%%%%%%%%%%%%%%%%%%%%%%%%%%%%%%%%%%%%%%%%%%%%%%%%%%%%%%%%%%%
When the curvature is small, it reduces to the Einstein-Hilbert
action with the cosmological constant
%**   4.21 %%%%%%%%%%%%%%%%%%%%%%%%%%%%%%%%%%%%%%%%%%%%%%%%%%%%%%%%%
\begin{eqnarray}
S_{eff} = \int d^D \xi \sqrt{-g} \left( \tilde{\lambda} 
+ \frac{1}{16 \pi G} R + O(R^2, \log \Lambda^2) \right),
\label{4.21}
\end{eqnarray}
%%%%%%%%%%%%%%%%%%%%%%%%%%%%%%%%%%%%%%%%%%%%%%%%%%%%%%%%%%%%%%%%%%%
with 
%**   4.22 %%%%%%%%%%%%%%%%%%%%%%%%%%%%%%%%%%%%%%%%%%%%%%%%%%%%%%%%%
\begin{eqnarray}
\tilde{\lambda} &=& \frac{D \Lambda^4}{8 (4 \pi)^2} +
\lambda, \nn\\
\frac{1}{16 \pi G} &=& \frac{D \Lambda^2}{24 (4 \pi)^2},
\label{4.22}
\end{eqnarray}
%%%%%%%%%%%%%%%%%%%%%%%%%%%%%%%%%%%%%%%%%%%%%%%%%%%%%%%%%%%%%%%%%%%
where we have introduced the momentum cutoff $\Lambda$ of the
Pauli-Villars type. Note that (\ref{4.22}) shows that we can choose
the effective cosmological constant $\tilde{\lambda}$ as small 
as we want, and the cutoff $\Lambda$ is of the order the Planck 
mass. It is quite interesting to ask why the topological action
has produced the Einstein-Hilbert action. This is because
the momentum cutoff $\Lambda$ breaks the topological symmetry
with keeping the general covariance. In other words, we have 
secretly introduced a seed for breaking the topological symmetry
by the form of the cutoff. Of course, it is an interesting
idea to make a conjecture that renormalization may induce such a
scale, but it seems to be quite difficult to prove this conjecture.

{}From this point of view, it is valuable to reconsider why 
the present formulation has produced the nontrivial matrix models
from the topological field theory. Originally, in membrane 
world, the matrix model has appeared to regularize the lightcone
supermembrane action with area-preserving diffeomorphisms where it 
has been remarkably shown that the action becomes exactly that 
of ten dimensional $SU(n)$ supersymmetric Yang-Mills theory 
reduced to $(0+1)$-dimensions \cite{DeWitt}. Similarly,
in our models, passing from the continuous topological
field theory to the discrete matrix model is equal to an
introduction of the regularization where the regularization
parameter corresponds to the size of the matrices. 
This type of the regularization breaks
only the topological symmetry, from which we can obtain the
nontrivial "dynamical" matrix models. It is very interesting
that the matrix model is equipped with such a natural
regularization scheme in itself. If the topological symmetry
is truly broken by some mechanism in order to make the 
topological field theory a physically vital theory, we believe
that theories equipped with some natural regularization scheme 
such as matrix model and induced gravity (pregeometry) would  
play an important role.

%%%%%%%%%%%%%%%%%%%%%%%%%%%%%%%%%%%%%%%%%%%%%%%%%%%%%%%%%%%%%%%%%%%%%
%%%%%%%%%%%%%%%%%%%%%%%%%%%%%%   SEC  5    %%%%%%%%%%%%%%%%%%%%%%%%%%
%%%%%%%%%%%%%%%%%%%%%%%%%%%%%%%%%%%%%%%%%%%%%%%%%%%%%%%%%%%%%%%%%%%%%
\section{ Background independent matrix models}

In this section, we shall construct two different kinds of
topological matrix models, which we call, Chern-Simons 
matrix model \cite{Smolin} and BF matrix model \cite{Oda3}.

Let us start with a background independent matrix model
which consists of only the hermitian matrices 
$X_\mu (\mu=0,1, \cdots, D-1)$. 
%**   5.1 %%%%%%%%%%%%%%%%%%%%%%%%%%%%%%%%%%%%%%%%%%%%%%%%%%%%%%%%%
\begin{eqnarray}
S_{CS}^D = \varepsilon^{\mu_1 \mu_2 \cdots \mu_D} 
Tr X_{\mu_1} X_{\mu_2} \cdots X_{\mu_D}.
\label{5.1}
\end{eqnarray}
%%%%%%%%%%%%%%%%%%%%%%%%%%%%%%%%%%%%%%%%%%%%%%%%%%%%%%%%%%%%%%%%%%%
Interestingly enough, we can construct such an action only in 
the case that $D$ is odd numbers since an action with even numbers
of $X_\mu$ is identically zero because of
the cyclic property of trace and the totally
antisymmetric property of the Levi-Civita tensor density. Thus
we will set $D$ to be $2d+1$ with $d \in {\bf Z_+} \cup \{0\}$
in this section.
Incidentally, the topological matrix model with any number
of $X_\mu$ will be built later.

The equations of motion derived from the action (\ref{5.1}) read
%**   5.3 %%%%%%%%%%%%%%%%%%%%%%%%%%%%%%%%%%%%%%%%%%%%%%%%%%%%%%%%%
\begin{eqnarray}
\varepsilon^{\mu \mu_1 \mu_2 \cdots \mu_{2d}} 
X_{\mu_1} X_{\mu_2} \cdots X_{\mu_{2d}} = 0.
\label{5.3}
\end{eqnarray}
%%%%%%%%%%%%%%%%%%%%%%%%%%%%%%%%%%%%%%%%%%%%%%%%%%%%%%%%%%%%%%%%%%% 
Note that (\ref{5.3}) does not include the background metric tensor
in comparison with the equations of motion derived from
IIB matrix models \cite{IKKT}. 
whose formal expression is provided by
%**   5.4 %%%%%%%%%%%%%%%%%%%%%%%%%%%%%%%%%%%%%%%%%%%%%%%%%%%%%%%%%
\begin{eqnarray}
\eta^{\mu\nu} \left[ X_\mu, \left[ X_\nu, X_\rho \right] \right]
= 0
\label{5.4}
\end{eqnarray}
%%%%%%%%%%%%%%%%%%%%%%%%%%%%%%%%%%%%%%%%%%%%%%%%%%%%%%%%%%%%%%%%%%% 
with the flat Minkowskian metric $\eta^{\mu\nu}$. 
At this stage, it is useful to find the classical solutions 
satisfying
the equations of motion (\ref{5.3}). One obvious solution is the
one satisfying the equation $\left[ X_\mu, X_\nu \right]
= 0$, that is, this solution has the form of the diagonal $N \times N$
matrix
%**   5.5 %%%%%%%%%%%%%%%%%%%%%%%%%%%%%%%%%%%%%%%%%%%%%%%%%%%%%%%%%
\begin{eqnarray}
 X_\mu =  \pmatrix{
X_\mu^{(1)} & {} & {} & {} \cr
{}          & {} & \ddots & {} \cr
{}          & {} &  {}    & X_\mu^{(N)} \cr
},
\label{5.5}
\end{eqnarray}
%%%%%%%%%%%%%%%%%%%%%%%%%%%%%%%%%%%%%%%%%%%%%%%%%%%%%%%%%%%%%%%%%%% 
which we call "classical space-time" in this paper.  Next nontrivial
solution is "string" solution given by
%**   5.6 %%%%%%%%%%%%%%%%%%%%%%%%%%%%%%%%%%%%%%%%%%%%%%%%%%%%%%%%%
\begin{eqnarray}
 X_\mu = \left( X_0, X_1, 0, \cdots, 0 \right),
\label{5.6}
\end{eqnarray}
%%%%%%%%%%%%%%%%%%%%%%%%%%%%%%%%%%%%%%%%%%%%%%%%%%%%%%%%%%%%%%%%%%% 
where we have considered the string along 1st axis without
losing generality. Similarly, "membrane" solution stretched out
in the direction of 1st and 2nd axes reads
%**   5.7 %%%%%%%%%%%%%%%%%%%%%%%%%%%%%%%%%%%%%%%%%%%%%%%%%%%%%%%%%
\begin{eqnarray}
 X_\mu = \left( X_0, X_1, X_2, 0, \cdots, 0 \right).
\label{5.7}
\end{eqnarray}
%%%%%%%%%%%%%%%%%%%%%%%%%%%%%%%%%%%%%%%%%%%%%%%%%%%%%%%%%%%%%%%%%%% 
It is obvious that this kind of solutions continues to exist until
"$(2d-1)$-brane"
%**   5.8 %%%%%%%%%%%%%%%%%%%%%%%%%%%%%%%%%%%%%%%%%%%%%%%%%%%%%%%%%
\begin{eqnarray}
 X_\mu = \left( X_0, X_1, X_2, \cdots, X_{2d-1}, 0 \right).
\label{5.8}
\end{eqnarray}
%%%%%%%%%%%%%%%%%%%%%%%%%%%%%%%%%%%%%%%%%%%%%%%%%%%%%%%%%%%%%%%%%%% 
Moreover, a solution associated with several "$k$-branes"
$(1 \le k \le 2d-1)$ can be built out of the above solution
for single "$k$-brane" in a perfectly similar way to
the case of IIB matrix model \cite{IKKT}.
For instance, the solution for two "strings" separated 
by the distance $b$ along 2nd axis is given by
%**   5.9 %%%%%%%%%%%%%%%%%%%%%%%%%%%%%%%%%%%%%%%%%%%%%%%%%%%%%%%%%
\begin{eqnarray}
X_0 = \pmatrix{
x_0 & 0   \cr
0   & x_0 \cr }, \ 
X_1 = \pmatrix{
x_1 & 0   \cr
0   & x_1 \cr }, \nn\\
X_2 = \pmatrix{
\frac{b}{2} & 0   \cr
0   & -\frac{b}{2} \cr }, \ 
X_3 = \cdots = X_{2d} = 0, 
\label{5.9}
\end{eqnarray}
%%%%%%%%%%%%%%%%%%%%%%%%%%%%%%%%%%%%%%%%%%%%%%%%%%%%%%%%%%%%%%%%%%% 
where $x_0$ and $x_1$ are certain nonzero elements.

Now let us turn our attention to the symmetries in the action
(\ref{5.1}). It is remarkable that as well as the conventional
gauge symmetry
%**   5.10 %%%%%%%%%%%%%%%%%%%%%%%%%%%%%%%%%%%%%%%%%%%%%%%%%%%%%%%%%
\begin{eqnarray}
 X_\mu \rightarrow X_\mu^\prime = U X_\mu U^{-1}
\label{5.10}
\end{eqnarray}
%%%%%%%%%%%%%%%%%%%%%%%%%%%%%%%%%%%%%%%%%%%%%%%%%%%%%%%%%%%%%%%%%%% 
with $U \in U(N)$, the action (\ref{5.1}) is invariant under
the local translation of the diagonal element
%**   5.11 %%%%%%%%%%%%%%%%%%%%%%%%%%%%%%%%%%%%%%%%%%%%%%%%%%%%%%%%%
\begin{eqnarray}
 X_\mu \rightarrow X_\mu^\prime =  X_\mu + V_\mu(X) \ {\bf 1}
\label{5.11}
\end{eqnarray}
%%%%%%%%%%%%%%%%%%%%%%%%%%%%%%%%%%%%%%%%%%%%%%%%%%%%%%%%%%%%%%%%%%% 
with $V_\mu(X)$ being not a matrix but a c-number function of $X_\mu$.
This symmetry is in sharp contrast with the matrix models 
\cite{M, IKKT} where $V_\mu$ is a global parameter of c-number.
Namely, the global translation in \cite{M, IKKT} is now
promoted to the local translation. In this respect,  
it is of interest to recall the following things. 
Firstly, in the matrix models
\cite{M, IKKT} the diagonal matrix like (\ref{5.5}) corresponds to  
the classical space-time coordinates while the non-diagonal
matrix describes the interactions. Hence the local symmetry
(\ref{5.11}) coincides with the local space-time translation 
at the classical level. Secondly, it is well known that general
relativity is the gauge theory with the local translation as
the gauge symmetry, so the existence of this symmetry might be
a signal of the existence of general relativity in this matrix
model though we need more studies to confirm this conjecture
in future. 

Thus far, we have considered the Chern-Simons
matrix model, but this model has some problems. In particular,
it is quite unsatisfactory that we cannot construct the 
matrix model in even space-time dimensions. Furthermore,
it seems to be difficult to make a supersymmetric extension of the
Chern-Simons matrix model without introducing the background
metric. Finally, it is at present unclear that the Chern-Simons matrix 
model has a relationship with general gravity. Luckily, we
have already met a similar situation to this in topological quantum
field theories where the Chern-Simons theory 
is replaced with the BF theory \cite{BF1, BF2, BF3, BF4, BF5} 
in order to overcome these impasse.
In the case of the matrix model at hand we also proceed with 
the same line of argument as the topological quantum 
field theories. 

Now we would like to present BF matrix model \cite{Oda3}
which has the form
%**   5.17 %%%%%%%%%%%%%%%%%%%%%%%%%%%%%%%%%%%%%%%%%%%%%%%%%%%%%%%%%
\begin{eqnarray}
S_n^D = \varepsilon^{\mu_1 \mu_2 \cdots \mu_D} 
Tr X_{\mu_1} X_{\mu_2} \cdots X_{\mu_n} B_{\mu_{n+1} \cdots \mu_D},
\label{5.17}
\end{eqnarray}
%%%%%%%%%%%%%%%%%%%%%%%%%%%%%%%%%%%%%%%%%%%%%%%%%%%%%%%%%%%%%%%%%%%
where a totally antisymmetric tensor matrix $B$ is introduced. 
In this respect let us recall that the original form of 
topological BF theory \cite{BF1, BF2, BF3, BF4, BF5} is
%**   5.18 %%%%%%%%%%%%%%%%%%%%%%%%%%%%%%%%%%%%%%%%%%%%%%%%%%%%%%%%%
\begin{eqnarray}
S_{BF} = \int \varepsilon^{\mu_1 \mu_2 \cdots \mu_D} 
Tr F_{\mu_1 \mu_2} \ B_{\mu_3 \cdots \mu_D},
\label{5.18}
\end{eqnarray}
%%%%%%%%%%%%%%%%%%%%%%%%%%%%%%%%%%%%%%%%%%%%%%%%%%%%%%%%%%%%%%%%%%% 
where the 2-form field strength $F$ is defined as $F = dA + A^2$.
Thus, precisely speaking, the straightforward generalization of the 
topological BF theory (\ref{5.18}) to the matrix model corresponds 
to the case of $n=2$ in (\ref{5.17}). Of course, owing to
the introduction of the matrix $B$ the action (\ref{5.17}) 
makes sense in arbitrary space-time dimension.

The classical equations of motion derived from the BF matrix model 
(\ref{5.17}) read
%**   5.19 %%%%%%%%%%%%%%%%%%%%%%%%%%%%%%%%%%%%%%%%%%%%%%%%%%%%%%%%%
\begin{eqnarray}
\varepsilon^{\mu_1 \mu_2 \cdots \mu_D} 
X_{\mu_1} X_{\mu_2} \cdots X_{\mu_n} = 0,
\label{5.19}
\end{eqnarray}
%%%%%%%%%%%%%%%%%%%%%%%%%%%%%%%%%%%%%%%%%%%%%%%%%%%%%%%%%%%%%%%%%%% 
%**   5.20 %%%%%%%%%%%%%%%%%%%%%%%%%%%%%%%%%%%%%%%%%%%%%%%%%%%%%%%%%
\begin{eqnarray}
\sum_{i=1}^n (-1)^{i-1} \varepsilon^{\mu \mu_1 \cdots 
\hat{\mu_i} \cdots \mu_D} X_{\mu_{i+1}} \cdots X_{\mu_n} 
B_{\mu_{n+1} \cdots \mu_D} X_{\mu_1} \cdots X_{\mu_{i-1}}
= 0,
\label{5.20}
\end{eqnarray}
%%%%%%%%%%%%%%%%%%%%%%%%%%%%%%%%%%%%%%%%%%%%%%%%%%%%%%%%%%%%%%%%%%% 
where $\hat{\mu_i}$ denotes that the index $\mu_i$ is excluded.
Note that apart from the number of $X_\mu$ , 
Eq.(\ref{5.19}) accords with (\ref{5.3}) in the Chern-Simons 
matrix theory. Thus the structure of the solutions with respect
to $X_\mu$ is almost the same as that case.
On the other hand, it is Eq.(\ref{5.20}) that appears for the 
first time in the BF matrix model. 
In fact, we can show that this equation has an important 
implication in relating the model at hand to general 
relativity \cite{Oda3}.

As for the gauge symmetries, besides the usual $U(N)$ gauge symmetry, 
at first glance the action (\ref{5.17}) looks like it might
be invariant under the following natural geralization of the local 
translation symmetry (\ref{5.11})
%**   5.21 %%%%%%%%%%%%%%%%%%%%%%%%%%%%%%%%%%%%%%%%%%%%%%%%%%%%%%%%%
\begin{eqnarray}
X_\mu &\rightarrow& X_\mu^\prime =  X_\mu + V_\mu(X) 
\ {\bf 1}, \nn\\
B_{\mu_{n+1} \cdots \mu_D} &\rightarrow& 
B_{\mu_{n+1} \cdots \mu_D}^\prime = B_{\mu_{n+1} \cdots \mu_D}
+ W_{\mu_{n+1} \cdots \mu_D}(X) \ {\bf 1}.
\label{5.21}
\end{eqnarray}
%%%%%%%%%%%%%%%%%%%%%%%%%%%%%%%%%%%%%%%%%%%%%%%%%%%%%%%%%%%%%%%%%%% 
However, it is interesting to notice that only the action 
(\ref{5.17}) with $n$ being even integers has such a local
translation symmetry while the action (\ref{5.17}) with odd $n$
has neither the local nor the global translation symmetry.

Now we will discuss some generalizations of the BF model
to include fermionic symmetry. Indeed, in the matrix
models \cite{M, IKKT} the fermionic symmetry, in
particular, the supersymmetry, was needed to guarantee
the cluster and BPS properties of instantons. 
One possibility is to add fermions of integer spin
to achieve a BRST-like symmetry. It is known that
the partition function of the BF theory is related
to the Ray-Singer torsion \cite{Schwarz2} 
while that of the BF theory with such a BRST-like 
symmetry correspondes to the Casson invariants.
We think that this statement
is valid even in the BF matrix model treated in 
this paper. Let us start by the following BRST-like
fermionic symmetry:
%**   5.23 %%%%%%%%%%%%%%%%%%%%%%%%%%%%%%%%%%%%%%%%%%%%%%%%%%%%%%%%%
\begin{eqnarray}
\delta X_\mu &=& \eta \psi_\mu, \ \delta \psi_\mu = 0, \nn\\
\delta \chi_{\mu_{n+1} \cdots \mu_D} &=& - \eta B_{\mu_{n+1} 
\cdots \mu_D}, \ \delta B_{\mu_{n+1} \cdots \mu_D} = 0.
\label{5.23}
\end{eqnarray}
%%%%%%%%%%%%%%%%%%%%%%%%%%%%%%%%%%%%%%%%%%%%%%%%%%%%%%%%%%%%%%%%%%% 
We can check explicitly the following action to be invariant
under the fermionic symmetry (\ref{5.23}):
%**   5.24 %%%%%%%%%%%%%%%%%%%%%%%%%%%%%%%%%%%%%%%%%%%%%%%%%%%%%%%%%
\begin{eqnarray}
S_n^D &=& \varepsilon^{\mu_1 \mu_2 \cdots \mu_D} 
Tr ( X_{\mu_1} X_{\mu_2} \cdots X_{\mu_n} B_{\mu_{n+1} 
\cdots \mu_D} \nn\\
& & {} - \sum_{i=1}^n X_{\mu_1} X_{\mu_2} \cdots X_{\mu_{i-1}}
\psi_{\mu_i} X_{\mu_{i+1}} \cdots X_{\mu_n} \chi_{\mu_{n+1}
\cdots \mu_D} )
\label{5.24}
\end{eqnarray}
%%%%%%%%%%%%%%%%%%%%%%%%%%%%%%%%%%%%%%%%%%%%%%%%%%%%%%%%%%%%%%%%%%%
{}For even integers $n$, this action is still invariant under 
the enlarged local translation which constitutes of Eq.(\ref{5.21}) 
and
%**   5.25 %%%%%%%%%%%%%%%%%%%%%%%%%%%%%%%%%%%%%%%%%%%%%%%%%%%%%%%%%
\begin{eqnarray}
\psi_\mu &\rightarrow& \psi_\mu^\prime =  \psi_\mu + v_\mu(X)
\ {\bf 1}, \nn\\
\chi_{\mu_{n+1} \cdots \mu_D} &\rightarrow& 
\chi_{\mu_{n+1} \cdots \mu_D}^\prime = \chi_{\mu_{n+1} \cdots \mu_D}
+ w_{\mu_{n+1} \cdots \mu_D}(X) \ {\bf 1}.
\label{5.25}
\end{eqnarray}
%%%%%%%%%%%%%%%%%%%%%%%%%%%%%%%%%%%%%%%%%%%%%%%%%%%%%%%%%%%%%%%%%%% 

A more interesting possibility of incorporaing fermions of half
integer spin would be to twist the action (\ref{5.24}) like the
topological quantum field theory \cite{Witten1}. Here note that 
even if the bosonic action
(\ref{5.1}) is nontrivial its BRST-like generalization (\ref{5.24})
is BRST-exact form so that we can use the twisting technique
developed in the reference \cite{Witten1}. 

%%%%%%%%%%%%%%%%%%%%%%%%%%%%%%%%%%%%%%%%%%%%%%%%%%%%%%%%%%%%%%%%%%%%%
%%%%%%%%%%%%%%%%%%%%%%%%%%%%%%   SEC  6    %%%%%%%%%%%%%%%%%%%%%%%%%%
%%%%%%%%%%%%%%%%%%%%%%%%%%%%%%%%%%%%%%%%%%%%%%%%%%%%%%%%%%%%%%%%%%%%%
\section{ Conclusions }

This work has explored various topological approaches that are
useful in constructing and understanding the matrix theory.
Making use of the Schild action explained in section 2, 
in sections 3 and 4 we have showed that
the space-time uncertainty principle of string and the matrix
models by Yoneya and IKKT are interpreted as being
induced from the spontaneous symmetry breakdown of the 
topological symmetry.

In section 5, we have presented an alternative form of topological
matrix models, which we have called the Chern-Simons and the BF
matrix models. This construction has a close parallel with
that of the conventional topological field theory which
is based on the continuous fields. These matrix
models are independent of the background metric, but give us
the nontrivial field equations such that the classical
solutions include the commuting space-time coordinates as well
as interesting p-brane solutions. We think it is worthwhile to
push forward line of inquiry of the models in more detail
in future. In particular, we would like to clarify how the 
$SL(2, Z)$ duality is realized in these matrix models by
following a similar procedure adopted in the reference
\cite{Oda4}.

\vs 1
%%%%%%%%%%%%%%%%%%%%%%%%%%%%%%%%%%%%%%%%%%%%%%%%%%%%%%%%%%%%%%%%%%
%%%%%%%%%%%%%%%%%%%%%%%% reference %%%%%%%%%%%%%%%%%%%%%%%%%%%%%%%
%%%%%%%%%%%%%%%%%%%%%%%%%%%%%%%%%%%%%%%%%%%%%%%%%%%%%%%%%%%%%%%%%%

\end{document}